\documentclass[12pt]{article}

\usepackage{epsf,macros,amsfonts,hyperref}
\usepackage{cite}
\newcommand\Zbar {{\bar{Z}}}
\newcommand\Zb {{\bar{Z}}}
\newcommand\yb {{\bar{y}}}
\begin{document}

\sloppy

\pagestyle{empty}

\hfill {\footnotesize hep-th/0209244}\\
\vspace{-0.3cm}

\hfill CRM-2867 (2002)\\

\addtolength{\baselineskip}{0.20\baselineskip}

\begin{center}

\vspace{26pt}
{\large \bf {BMN Correlators by Loop Equations}}
\end{center}
\vspace{26pt}

\begin{center}
B.\ Eynard\hspace*{0.05cm}\footnote{ E-mail: eynard@spht.saclay.cea.fr, 
Permanent address: Service de Physique Th\'{e}orique de Saclay,
F-91191 Gif-sur-Yvette Cedex, France. }
\\
\vspace{6pt}
{\sl Centre de recherches math\'{e}matiques, Universit\'{e} de Montr\'{e}al}
\\
{\sl C. P. 6128, succ.\ centre ville, Montr\'eal, Qu\'ebec, Canada H3C 3J7}
\vspace{26pt}

{\sl C.\ Kristjansen}\hspace*{0.05cm}\footnote{ E-mail: 
kristjan@alf.nbi.dk, work supported by the Danish Natural Science Research 
Council}\\
\vspace{6pt}
{\sl The Niels Bohr Institute,} \\
{\sl Blegdamsvej 17, DK-2100 Copenhagen \O}\\

\end{center}

\vspace{20pt}
\begin{center}
{\bf Abstract}
\end{center}
In the BMN approach to ${\cal N}=4$ SYM a large class of correlators
of interest are expressible
in terms of expectation values of traces of
words in a zero-dimensional Gaussian complex 
matrix model. We develop a loop-equation based, analytic strategy for 
evaluating such expectation values to any order in the genus
expansion. We reproduce the expectation values which were needed for
the calculation of the one-loop, genus one correction to the 
anomalous dimension
of BMN-operators and which were earlier obtained by combinatorial
means. Furthermore, we present the expectation values needed for the
calculation of the one-loop, genus two correction.


\newpage

\pagestyle{plain}
\setcounter{page}{1}
\section{Introduction}
Recent progress in string and gauge theory~\cite{Blau:2002dy,
Metsaev:2002re,Berenstein:2002jq} has brought to light an interesting 
pp-wave/BMN-correspondence which is a special version of the celebrated
AdS/CFT correspondence. The pp-wave is a ten-dimensional geometry
which can be obtained as a Penrose limit of AdS$_5\times$S$_5$ and 
which constitutes a background where
it is possible to quantize type IIB string theory in light cone
gauge~\cite{Blau:2002dy,Metsaev:2002re,Metsaev:2001bj}.
 BMN stands for Berenstein, Maldacena and Nastase who identified
the gauge theory dual as a special sector of ${\cal N}=4$ SYM based
on gauge group SU(N) with a 
certain limit understood.

The BMN sector of ${\cal N}=4$ SYM consists of operators which carry a
large R-charge, $J$, associated with a selected SO(2) sub-group of the
full SO(6) R-symmetry group and for which $\Delta-J$ is finite where
$\Delta$ is the conformal dimension. It has been argued that the
quantum corrections to correlation functions involving such operators
only depend on $g_{YM}$ via the parameter $\lambda'=(g_{YM}^2N)/J^2$
\cite{Berenstein:2002jq,Gross:2002su,Santambrogio:2002sb} and the BMN
limit is given by the scaling prescription
\begin{equation}
g_{YM}\hspace{0.3cm}\mbox{fixed,}\hspace{0.3cm}
J,N\rightarrow \infty\hspace{0.3cm}\mbox{with}\hspace{0.3cm}
g_2=\frac{J^2}{N} \mbox{ fixed}
\end{equation}
which in particular renders $\lambda'$ finite.\footnote{It appears 
that this limit is the same for gauge groups
SU(N) and U(N). In this paper we shall be considering
gauge group U(N).} As shown in~\cite{Kristjansen:2002bb,Berenstein:2002sa,
Constable:2002hw} despite being a large-N limit the BMN limit is not
a planar limit. Diagrams of all genera survive the limit and
contributions of genus $h$ are weighted by a factor $(g_2)^{2h}$. One
could say that the BMN approach to ${\cal N}=4$ SYM introduces a new
't Hooft expansion with a new   gauge coupling constant $\lambda'$
and a new genus counting parameter, $g_2$. However, the BMN {\it limit}
is not a new 't Hooft {\it limit} because the genus counting parameter
remains finite as the limit is taken. Rather, the BMN limit is an
interesting new double scaling limit much like the one encountered in
the study of 2D quantum 
gravity~\cite{Brezin:1990rb,Douglas:1990ve,Gross:1990vs}. 

 To introduce the R-charge
of the BMN approach we single out two of the  six scalars 
$\phi_i(x)$, $i=\{1,\ldots,6\}$, which
transform under the SO(6) R-symmetry group, say $\phi_5$
and $\phi_6$, and form the complex combination
\begin{equation}
Z(x)=\frac{1}{\sqrt{2}}\left(\phi_5(x)+i\phi_6(x)\right)
\label{Zdef}
\end{equation}
Then we define the R-charge, J, as the quantum number conjugate to the
phase of $Z$. As mentioned above, operators which survive the BMN
limit are characterized by having $J$ very large and $\Delta-J$
finite. In practice this means that such operators contain a large
number of $Z$-fields and a finite number of impurities in the form
of fields not carrying R-charge such as $\phi_1$, $\phi_2$, $\phi_3$
and $\phi_4$. In ${\cal N}=4$ SYM and in particular in its BMN sector
the space-time dependence of two- and three-point functions is fixed
by conformal invariance. At the classical level the 
calculation of such correlators then reduces to the calculation of 
expectation values
in a zero dimensional Gaussian complex matrix model. For protected
operators this statement trivially remains true when interactions are
included and for non-protected operators a similar simplification can
be obtained even at the quantum level if one introduces effective
vertices~\cite{Kristjansen:2002bb,Constable:2002hw}. (This procedure
has so far only been implemented at one-loop level.) 
In reference\cite{Verlinde:2002ig}
it was proposed that only two-point functions of appropriately defined
multi-trace operators would have a string theory interpretation and 
this point of view has been supported by gauge theory 
calculations~\cite{Beisert:2002bb,Gross:2002mh}. This implies
 that extracting 
information about pp-wave strings from the gauge-theory
reduces to  determining the expectation value of traces of
words in a zero-dimensional Gaussian complex matrix model.
So far genuine matrix model techniques have only been exploited in
the calculation of a very limited set of expectation 
values~\cite{Kristjansen:2002bb,Constable:2002hw} whereas
the major part of those obtained were determined by combinatorial means.
For higher genera
combinatorial arguments become very involved. From the string theory
point of view higher genera contributions are most interesting because
they encode information about string interactions. So far gauge
theory calculations were only pursued up to and including genus one.

In the present paper we shall develop a loop-equation based, analytic
strategy which allows us to calculate by recursion expectation values
of products of arbitrary traces of 
words in a Gaussian complex one-matrix model to any order
in the genus expansion. 
The outline of our paper is the
following. First, in section~2 we explain in more detail how to
reduce the calculation of two-point functions in ${\cal N}=4$ SYM
to the calculation of matrix model expectation values, focusing on
the two-point function of the so-called BMN operators. Next, in section~3
we introduce the notation necessary for our matrix model investigations
and list the matrix model expectation values which are needed to find
respectively the genus one and the genus two, one-loop correction
to the anomalous dimension of the BMN-operators.
In section~4
we derive the two basic relations on which all our
considerations are based; the split and merge rule respectively. 
As a first application of these rules we reproduce in
section~5 all the matrix model expectation values needed for the above
mentioned genus one calculation by purely analytic
computations. Subsequently, in section~6 we determine the correlators
needed for the genus 2 calculation and finally in section~7 we show
how our strategy allows us to find the expectation value of 
traces of arbitrary
words to any order in the genus expansion. Section~8 is devoted to 
correlators which can be calculated exactly and section~9
contains our conclusions.

\vspace*{0.3cm}
\noindent
{\bf Note:} As we were completing our manuscript a related, interesting 
paper appeared where another loop equation based technique is applied to
the study of the (planar) BMN-limit~\cite{Koch:2002nq}.

\section{From ${\cal N}=4$ SYM to matrix model}

The field content of ${\cal N}=4$ SYM in four dimensions consists
of the scalars
$\phi_i(x)$, $i\in\{1,\ldots,6\}$, a space-time vector $A_{\mu}(x)$ and a
sixteen component spinor $\psi(x)$. These fields are Hermitian $N\times N$
matrices and can be expanded in terms of the generators $T^a$ of the gauge
group $U(N)$, for instance

\begin{equation}
(\phi_i)_{\alpha\beta}(x)=\sum_{a=0}^{N^2-1} \phi^a_i(x)T^a_{\alpha\beta}
\end{equation}
The generators are normalized as follows
\begin{equation}
\tr [T^a,T^b]=\delta^{ab},\hspace{0.7cm}
\sum_{a=0}^{N^2-1}T^a_{\alpha\beta}T^a_{\gamma\delta}=
\delta_{\delta\alpha}\delta_{\beta\gamma}
\end{equation}
and the Euclidean action reads
\begin{eqnarray}
\lefteqn{\hspace{-3.0cm}S=\frac{2}{g_{YM}^2}\int d^4x \tr\left(
\frac{1}{4}F_{\mu\nu}F_{\mu\nu}+\frac{1}{2}D_{\mu}\phi_iD_{\mu}\phi_i
-\frac{1}{4}[\phi_i,\phi_j][\phi_i,\phi_j]\right.} \nonumber \\ 
&& \left.+\frac{1}{2}\bar{\psi}\Gamma_{\mu}D_{\mu}\psi-
\frac{i}{2}\bar{\psi}\,\Gamma_i[\phi_i,\psi]\right)\label{SYM}
\end{eqnarray}
where 
$F_{\mu\nu}=\partial_\mu A_{\nu}-\partial_{\nu}A_{\mu}-i[A_{\mu},A_{\nu}]$
and the covariant derivative is $D_{\mu}\phi_i=\partial_{\mu}
\phi_i-i[A_{\mu},\phi_i]$. Furthermore,
$(\Gamma_{\mu},\Gamma_i)$ are the ten-dimensional Dirac matrices in the 
Majorana-Weyl representation.
Working in Feynman gauge, the propagators of the scalar fields take the form
\begin{equation}
\langle (\phi_i)_{\alpha\beta}(x)(\phi_j)_{\gamma\delta}(0)\rangle=
\frac{g_{YM}^2}{8 \pi^2 x^2}\,\delta_{ij}
\delta_{\alpha\delta}\delta_{\beta\gamma} \label{Feyn1}
\end{equation}
and in particular (cf.\ eqn.~\rf{Zdef})
\begin{equation}
\langle {\bar Z}_{\alpha\beta}(x)Z_{\gamma\delta}(0)\rangle=
\frac{g_{YM}^2}{8 \pi^2 x^2}\,
\delta_{\alpha\delta}\delta_{\beta\gamma} \label{Feyn2}
\end{equation}
\beq\label{ZZ}
\langle {\bar Z}_{\alpha\beta}(x)\Zb_{\gamma\delta}(0)\rangle
=\langle  Z_{\alpha\beta}(x)Z_{\gamma\delta}(0)\rangle
=0
\eeq
 Operators ${\cal O}(x)$ which belong to the BMN sector of ${\cal N}=4$
SYM are characterized by containing a large number of $Z$-fields and a finite
number of impurities in the form of fields not carrying R-charge. As an 
example, let us consider the most studied, so-called BMN-operator
\begin{equation}\label{BMNop}
{\cal O}_{12,n}^J(x)\equiv \frac{1}{\sqrt{N^{J+2}J}}
\sum_{p=0}^J e^{2\pi i p n/J}
\tr\left(\phi_1(x)Z^p(x)\phi_2(x)Z^{J-p}(x)\right)
\end{equation}
From the Feynman rules~\rf{Feyn1}, \rf{Feyn2} and~\rf{ZZ}, 
(or alternatively from
conformal invariance) it follows that the tree level two-point function
of BMN-operators can be written as
$$
\langle {\cal O}_{12,n}^J(x)\bar{{\cal O}}_{12,m}^J(0)\rangle
=\left(\frac{g_{YM}^2}{8\pi^2 x^2}\right)^{J+2}
\sum_{p,q=0}^J e^{2\pi i(np-mq)/J}
\langle\tr(\phi_1 Z^p \phi_2 Z^{J-p})
 \tr(\phi_1 \bar{Z}^{J-q}\phi_2\bar{Z}^q)
\rangle 
$$
where the space-time independent matrix valued fields, $\phi_i$ and $Z$
should be contracted using the following Feynman rules
\begin{eqnarray}
\langle (\phi_i)_{\alpha\beta}(\phi_j)_{\gamma\delta}\rangle&=&
\delta_{ij}
\delta_{\alpha\delta}\delta_{\beta\gamma} \label{matrix1} \\
\langle {\bar Z}_{\alpha\beta}Z_{\gamma\delta}\rangle&=&
\delta_{\alpha\delta}\delta_{\beta\gamma} \label{matrix2} 
\end{eqnarray}
The contraction of the $\phi$-fields can easily be done by hand and we
are left with
\begin{equation}
\langle {\cal O}_{12,n}^J(x)\bar{{\cal O}}_{12,m}^J(0)\rangle
=\left(\frac{g_{YM}^2}{8\pi^2 x^2}\right)^{J+2}
\sum_{p,q=0}^J e^{2\pi i(np-mq)/J}
\langle \tr(Z^{J-p} \bar{Z}^{J-q})
\tr(Z^p \bar{Z}^q)\rangle 
\label{twopoint}
\end{equation}
Now, the remaining expectation value can be identified as an expectation
value in a zero dimensional Gaussian complex matrix model, namely
\begin{equation}
\langle \tr(Z^{J-p} \bar{Z}^{J-q})
\tr(Z^p \bar{Z}^q)\rangle=
\int dZ d\bar{Z} \exp\left(-\tr(\bar{Z}Z)\right) \tr(Z^{J-p} \bar{Z}^{J-q})
\tr(Z^p \bar{Z}^q)
\label{identification}
\end{equation}
Here $dZ d\bar{Z}$ is defined as
\begin{equation}
dZ d\bar{Z} =\prod_{i,j=1}^N \frac{d\Re Z_{ij} d \Im Z_{ij}}{\pi}
\label{measure}
\end{equation}
such that $\int dZ d\bar{Z} e^{-\tr(\bar{Z}Z)}=1$. The 
identification~\rf{identification} holds because the matrix model 
measure~\rf{measure} combined with the Gaussian action precisely 
give rise to the contraction
rule~\rf{matrix2}. The action and the measure carry a $U(N)\times U(N)$
symmetry corresponding to the transformation $Z\rightarrow UZV^{\dagger}$
with $U$ and $V$ unitary. Expectation values of operators likewise carrying
this symmetry, i.e. traces of products of $(\bar{Z}Z)$ can be calculated
even for arbitrary $U(N)\times U(N)$ invariant potential
order by order in the genus expansion using 
loop equations~\cite{Ambjorn:1992xu}. Expectation values of operators
consisting of products of traces involving only $Z$'s or $\bar{Z}$'s
can likewise be obtained by well-established methods, namely by
character expansion~\cite{Kostov:1996bs,Kostov:1997bn} or by the method
of Ginibre~\cite{Ginibre}. Notice that the object appearing
in~\rf{identification} does not belong to either of these classes of
correlators. The aim of the present paper is to develop a method which 
allows us to deal with general correlators composed of traces of
arbitrary words of $Z$ and $\bar{Z}$. 

It is obvious that any tree-level two point function of operators in
the BMN sector of ${\cal N}=4$ SYM can be reduced in the above manner.
One pulls out the space-time factor, contract by hand the {\it
finite} number of impurities and one is left with a matrix model
expectation value. By making use of so-called effective vertices one
can also reduce one-loop corrections to two-point functions to matrix
model expectation values~\cite{Kristjansen:2002bb,Constable:2002hw}.

As explained in the introduction correlation functions in the BMN
sector of ${\cal N}=4$ SYM have an expansion in powers of
$\frac{J^4}{N^2}$,
the genus counting parameter, and we are interested in determining (at
least) the first terms in this expansion. For that purpose it is
convenient to decompose our matrix model expectation values into
connected and disconnected parts, f.\ inst.\
$$
\langle \tr(Z^p \bar{Z}^q)\tr (Z^{J-p}\bar{Z}^{J-q})\rangle=
\langle\tr(Z^p \bar{Z}^q)\rangle \langle \tr
(Z^{J-p}\bar{Z}^{J-q})\rangle
+\langle \tr(Z^p \bar{Z}^q)\tr (Z^{J-p}\bar{Z}^{J-q})\rangle_{conn}
$$
as the connected part is down by a factor of $\frac{1}{N^2}$ compared to
the disconnected one.

Furthermore, it is convenient to work with generating functionals for
expectation values in stead of working with the expectation values 
themselves. For instance, let us define
\begin{equation}
W_{1,1}(x_1,y_1;x_2,y_2)=\left<
\tr\left( \frac{1}{x_1-Z}\frac{1}{\bar{y}_1-\bar{Z}}\right)
\tr\left( \frac{1}{x_2-Z}\frac{1}{\bar{y}_2-\bar{Z}}\right)\right>_{conn}
\end{equation}
where $x_1$, $y_1$, $x_2$, $y_2$ are to be viewed as auxiliary
variables. Then we have 
\begin{eqnarray}\label{mixmatrix}
\lefteqn{\
W_{1,1}(X\ee{\frac{-i\pi n}{J}},X\ee{\frac{-i\pi m}{J}};
X\ee{\frac{i\pi n}{J}},
X\ee{\frac{i\pi m}{J}}) = }\\
&& e^{i \pi (m-n)}\,
\sum_{J=0}^\infty (X\bar{X})^{-J-2} \sum_{p,q=0}^J 
\left< \tr (Z^{J-p} \Zbar^{J-q}) \tr (Z^p \Zbar^q) \right>_{conn}
\ee{2 i\pi(n p -m q)/J} \nonumber
\end{eqnarray}
which immediately allows us to extract the sum appearing in~\rf{twopoint}
also known as the tree-level mixing matrix. So far, the tree-level
mixing
matrix has been calculated to order $\frac{J^4}{N^2}$ (genus one)
in~\cite{Kristjansen:2002bb,Constable:2002hw} and to order 
$\left(\frac{J^4}{N^2}\right)^2$ in~\cite{Constable:2002hw}. Furthermore, 
the one-loop correction to the two-point function was calculated to
genus one in~\cite{Kristjansen:2002bb,Constable:2002hw}. In the
following section we list the matrix model expectation values or
rather the generating functions needed for that computation. We
likewise list the ones needed to extend that calculation
to genus two. Later we shall determine all of these functions..

\section{Definitions and Notation}

We consider a complex Gaussian matrix model whose partition function
is given by
\beq
{\cal Z} = \int d\mu\, \ee{-S}=\int \D{Z}\D{\bar{Z}} \,\, \ee{-N\tr {Z\bar{Z}} } 
\eeq
where the integration runs over complex $N\times N$ matrices.
Note that there appears a factor of $N$ in front of the
action. This factor is introduced only for convenience and can easily
be scaled away in the final results. Let us define the following
generating functionals, also denoted as loop functions.
 \beq
\omega(x) = {1\over N} \left< \tr {1\over x-Z} \right> = {1\over x}
\virg
\bar\omega(y) = {1\over N} \left< \tr {1\over \yb-\Zbar} \right> = {1\over \yb}
\label{omega}
\eeq

\beq
W_1(x,y) ={1\over N} \left< \tr {1\over x-Z} {1\over \yb-\Zbar}
\right> 
\label{W}
\eeq

\beq
W_2(x,y,x',y') = {1\over N} \left< \tr {1\over x-Z} {1\over \yb-\Zbar}
{1\over x'-Z} {1\over \bar{y}'-\Zbar} \right> 
\label{W2}
\eeq

\beq
W_{1,1}(x,y;x',y') = \left< \tr {1\over x-Z} {1\over \bar{y}-\Zbar} \tr
{1\over x'-Z} {1\over \yb'-\Zbar} \right>_{\rm conn}
\label{W11}
\eeq

\beq
U_1(x;x',y') = \left< \tr {1\over x-Z} \tr {1\over x'-Z} {1\over \yb'-\Zbar}
\right>_{\rm conn}
\label{U1}
\eeq
We have normalized these functions so that their leading term
in the large-N expansion is of order one. Knowing the leading
order contributions to these functions for large N as well as
the next to leading
order contribution to $W_1(x,y)$
suffices for the
calculation of the one-loop, genus one correction to the anomalous
dimension of the BMN operators. However, we shall be interested in 
more general loop functions. We define
\begin{eqnarray}\label{genusexp1}
\lefteqn{W_{l_1,\dots,l_n}(x_{1,1},y_{1,1},\dots,x_{1,l_1},y_{1,l_1};\dots;x_{n,1},y_{n,1},\dots,x_{n,l_n},y_{n,l_n})  }\\
&&= N^{n-2} \left< \prod_{j=1}^{n} \tr \left( \prod_{i=1}^{l_j} {1\over
x_{j,i}-Z}{1\over \yb_{j,i}-\Zb} \right) \right>_{\rm conn}\nonumber \\
&&=\sum_{h=0}^{\infty} \frac{1}{N^{2h}}
W_{l_1,\dots,l_n}^{(h)}
(x_{1,1},y_{1,1},\dots,x_{1,l_1},y_{1,l_1};\dots;x_{n,1},y_{n,1},
\dots,x_{n,l_n},y_{n,l_n})\nonumber 
\end{eqnarray}
This function is invariant under permutation of the various traces,
under cyclic permutation of the factors inside a given trace and it is
changed to its complex conjugate under $x \leftrightarrow y$. We can
represent it with a Young diagram like graph as follows with $l_1\leq l_2
\ldots\leq l_n$
\begin{figure}[h]
\begin{center}
\qquad\epsfbox{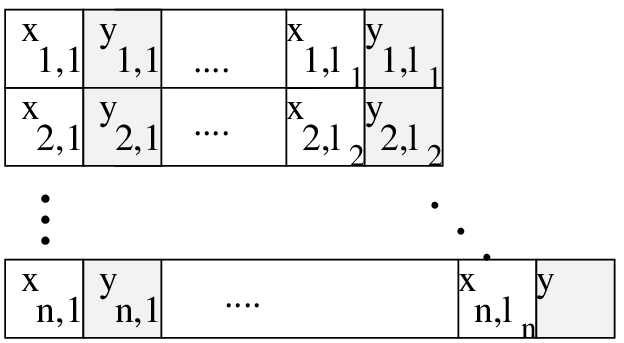} 
\end{center}
\end{figure}
\vspace*{-0.65cm}

\noindent
We also define
\begin{eqnarray}\label{genusexp2}
\lefteqn{
U_{l_1,\dots,l_n}(x;x_{1,1},y_{1,1},\dots,x_{1,l_1},
y_{1,l_1};\dots;x_{n,1},y_{n,1},\dots,x_{n,l_n},y_{n,l_n}) 
 } \\
& & =N^{n-1} \left< \tr{1\over x-Z}\prod_{j=1}^{n} 
\tr \left( \prod_{i=1}^{l_j}
{1\over x_{j,i}-Z}{1\over \yb_{j,i}-\Zb} \right) \right>_{\rm conn} 
\nonumber \\
&& =\sum_{h=0}^{\infty}
\frac{1}{N^{2h}}U_{l_1,\dots,l_n}^{(h)}(x;x_{1,1},y_{1,1},\dots,x_{1,l_1},
y_{1,l_1};\dots;x_{n,1},y_{n,1},\dots,x_{n,l_n},y_{n,l_n})\nonumber
\end{eqnarray}
which we can similarly represent with a Young diagram like graph

\begin{figure}[h]
\begin{center}
\qquad\epsfbox{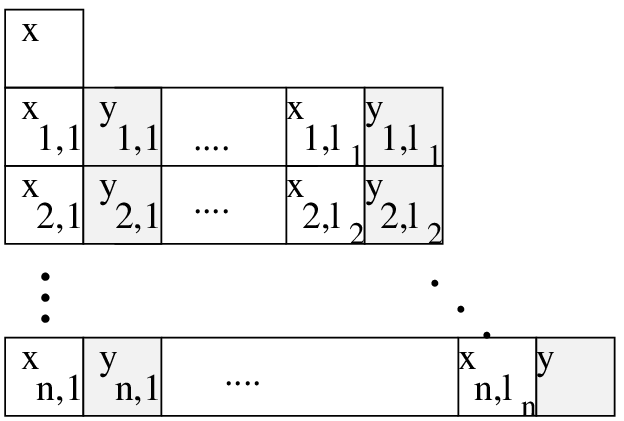} 
\end{center}
\end{figure}
\vspace*{-0.55cm}

\noindent
where $l_1\leq l_2\ldots \leq l_n$.
Calculating the one-loop, {\it genus two} correction to the two-point
function of BMN-operators would require the knowledge of the third order 
contribution to $W_1(x,y)$,
the next to
leading order contribution to the functions~\rf{W2}--\rf{U1} as well as
the leading order contribution to the functions
$U_2(x;x_1,y_1,x_2,y_2)$, $W_3(x_1,y_1,x_2,y_2,x_3,y_3)$, 
$U_{1,1}(x;x_1,y_1;x'_1,y'_1)$ and
$W_{1,2}(x,y;x_1,y_1,x_2,y_2)$.
In section~6 we shall show how to determine these and in section~7 we
shall describe a general strategy for determining any multi-loop
function
or equivalently any expectation value of traces of words to any order
in the genus expansion.

\section{Loop Equations}
All our computations will be based on two simple rules which can be
derived by loop equation techniques, based on the fact that the
matrix
model partition function is invariant under field 
redefinitions~\cite{Makeenko:pb}.
Here we restrict ourselves to considering the case of a 
complex matrix model with a Gaussian
potential which is the case of interest for the BMN sector of ${\cal
N}=4$ SYM. This is, however, just an almost trivial application of
a method which works under much more general circumstances and will
be presented for a Hermitian two-matrix model with arbitrary $U(N)$ 
invariant potentials in~\cite{Bertrand}.

\subsection{Split Rule}
Consider the field redefinition
\beq
Z \to Z + \epsilon A {1\over x-Z} B 
\label{redef}
\eeq
 This redefinition gives rise to
the following change of the measure
\beq\label{dS}
\delta\left(d Z d \bar{Z}\right)=
2 \Re\left(\epsilon\, \tr A{1\over x-Z} \tr{1\over x-Z}B\right) dZ d\bar{Z} 
+O(\epsilon^2)
\eeq
If $A$ or $B$ depends on $Z$ or $\bar{Z}$ there will be additional
contributions which are obtained by applying the usual chain rule 
in combination with the split and merge rules.
Obviously, under~\rf{redef} the action changes as
\beq\label{dm}
\delta S=2 N \Re \left(\epsilon\, \tr\left(\bar{Z}
A\frac{1}{x-Z}B\right)\right)
\eeq
The relations~\rf{dS} and~\rf{dm} hold for arbitrary complex $\epsilon$,
in particular for $\epsilon$ purely real or purely imaginary. Therefore
we conclude
\beq\label{split}
\left< \tr A\frac{1}{x-Z}\tr \frac{1}{x-Z}B\right> 
=N \left<\tr \bar{Z}A\frac{1}{x-Z}B\right>
\eeq

\subsection{Merge Rule}
Here we consider the following field redefinition
\beq
Z \to Z + \epsilon A \tr {1\over x-Z} B
\eeq
for which the change in the measure is
\beq
\delta\left(d Z d \bar{Z}\right)=
2 \Re \left(\epsilon \tr A{1\over x-Z} B {1\over x-Z}\right) d Z d
\bar{Z}+O(\epsilon^2)
\eeq
Again, if $A$ or $B$ depends on $Z$ or $\bar{Z}$ there will be additional
contributions which are obtained by applying the usual chain rule 
in combination with the split and merge rules.
The change of the action is obvious and our final merge rule reads
\beq\label{merge}
\left< \tr \left( A{1\over x-Z} B {1\over x-Z}\right) \right>=
N \left< \tr \bar{Z}A \,\,\tr\frac{1}{x-Z}B \right> 
\eeq

\section{Functionals needed for the one-loop, genus one computation} 
 In this section we shall determine the leading order contribution
for large $N$ to the loop-functions~\rf{W}--\rf{U1} as well
as the next to leading order contribution to~\rf{W}. As mentioned
above these are the objects needed for the computation of
the one-loop, genus one correction to the anomalous dimension of the
BMN-operators. In section~8 we will show that the functionals
\rf{W} and~\rf{U1} can be calculated exactly but to expose the completeness
of our loop equation method we shall derive the leading order contributions
to these below as well.

\subsection{$\omega(x)$ and $W_1(x,y)$ to leading order}

Considering the field redefinition $\delta \Zb = {1\over x-Z}$ we easily
get

\beq
0 = -1+ x\omega(x) 
\eeq
which is true to all orders in $\frac{1}{N^2}$ and gives
\beq
\omega(x)=\frac{1}{x}
\eeq
This result of course trivially follows from symmetry arguments.
Next, we make use of the field redefinition
$\delta Z = {1\over x-Z}{1\over \yb-\Zb}$ and obtain

\begin{figure}[h]
\begin{center}
\qquad\epsfbox{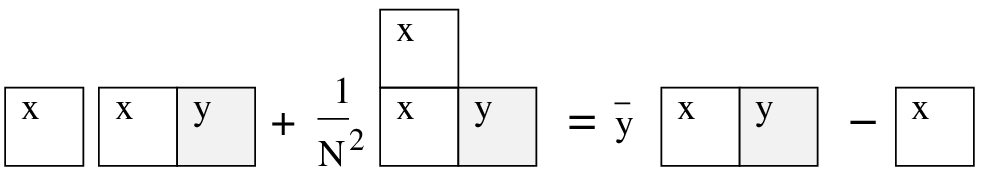} 
\end{center}
\end{figure}
\vspace*{-0.7cm}

\beq\label{W1loop}
\omega(x)W_1(x,y)+\frac{1}{N^2} U_1(x;x,y)=\yb W_1(x,y)-\omega(x)
\eeq

\noindent
Above, a space between two Young diagrams signifies multiplication of the
corresponding functions.
To leading order in $\frac{1}{N^2}$ we can neglect the second term 
in~\rf{W1loop} and we
get
\beq\label{W10}
W_1^{(0)}(x,y) = {1\over x\yb-1} 
\eeq
which is in accordance with the simple combinatorial result
\beq
{1\over N} \left<\tr Z^{J} \Zb^{J} \right> 
= 1  + O({1}/{N^2})
\eeq

\subsection{$W_2$, $U_1$ and $W_{1,1}$ to leading order}
Performing the change of variable 
$\delta{Z} = {1\over x_1-Z}{1\over \yb_1-\bar{Z}}{1\over x_2-Z}
{1\over \yb_2-\bar{Z}}$ leads to: \newpage
\vspace*{-0.2cm}
\begin{figure}[h]
\begin{center}
\centerline{{\epsfxsize=14.0cm\epsfbox{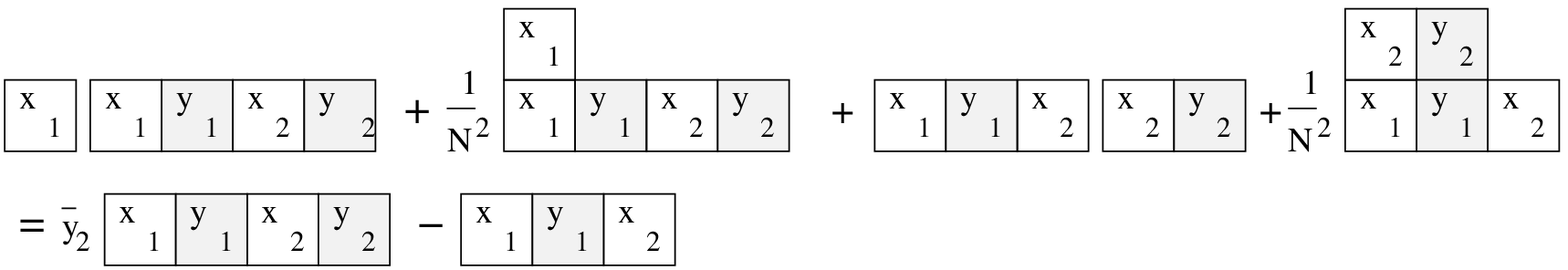}}}
\end{center}
\vspace*{-1.7cm}
\end{figure}

\bea\label{W2loop}
\lefteqn{ W_2(x_1,y_1,x_2,y_2)(\yb_2-\frac{1}{x_1})}\\
&&=\left(1+W_1(x_2,y_2)\right) 
{W_1(x_1,y_1)-W_1(x_2,y_1)\over x_2-x_1} \nonumber \\
&&{}+ {1\over N^2}U_2(x_1;x_1,y_1,x_2,y_2) + {1\over N^2} 
{W_{1,1}(x_1,y_1;x_2,y_2)-W_{1,1}(x_2,y_1;x_2,y_2)\over x_2-x_1}
\nonumber
\eea

\noindent
In the equation above we have used fractional decomposition to express the 
quantities represented by the two last Young diagrams in each line in
terms of usual $W$ functions with an even number of arguments. 
From equation~\rf{W2loop} we can easily find the leading contribution to
$W_2(x_1,y_1,x_2,y_2)$ for large $N$, namely
\beq\label{W2leading}
W_2^{(0)}(x_1,y_1,x_2,y_2) =  {x_1\yb_1 x_2\yb_2 
\over (x_1\yb_1-1) (x_2\yb_1-1)(x_1\yb_2-1)(x_2\yb_2-1)}  
\eeq
which reproduces the combinatorial result
\beq
{1\over N} \left<\tr Z^{J-p} \Zb^{J-q} Z^p \Zb^q \right> 
=1+{\rm Min}\left[p,q,J-p,J-q\right]  + O({1}/{N^2})
\eeq
Next, carrying out the change of variables 
$\delta \Zb = {1\over x-Z}\tr{1\over x_1-Z}{1\over \yb_1-\Zb}$ we get the
following simple relation

\begin{figure}[h]
\begin{center}
\qquad\epsfbox{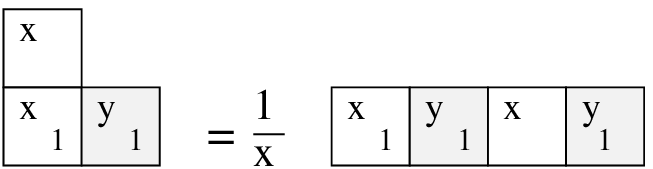} 
\end{center}
\end{figure}
\vspace{-.7cm}
\beq\label{loopeqU1}
 U_1(x;x_1,y_1) = \frac{1}{x}W_2(x_1,y_1,x,y_1)
\eeq
i.e.\ to leading order
\beq\label{U10}
U_1^{(0)}(x;x_1,y_1)=\frac{x_1 \yb_1^2}{(x_1\yb_1-1)^2(x\yb_1-1)^2}
\eeq
which is in agreement with the combinatorial result
\beq
\left<\tr \Zb^{p} \tr Z^{J} \Zb^{J-p} \right>_{\rm conn} = p(J-p+1) + O(1/N^2)
\eeq
Finally, we consider the field redefinition
$\delta{Z} = {1\over x_1-Z}{1\over \yb_1-\bar{Z}} \tr
{1\over x_2-Z}{1\over \yb_2-\bar{Z}}$ which leads to the following
equation for $W_{1,1}(x_1,y_1;x_2,y_2)$
\begin{figure}[h]
\begin{center}
\centerline{\epsfxsize=14.0cm \epsfbox{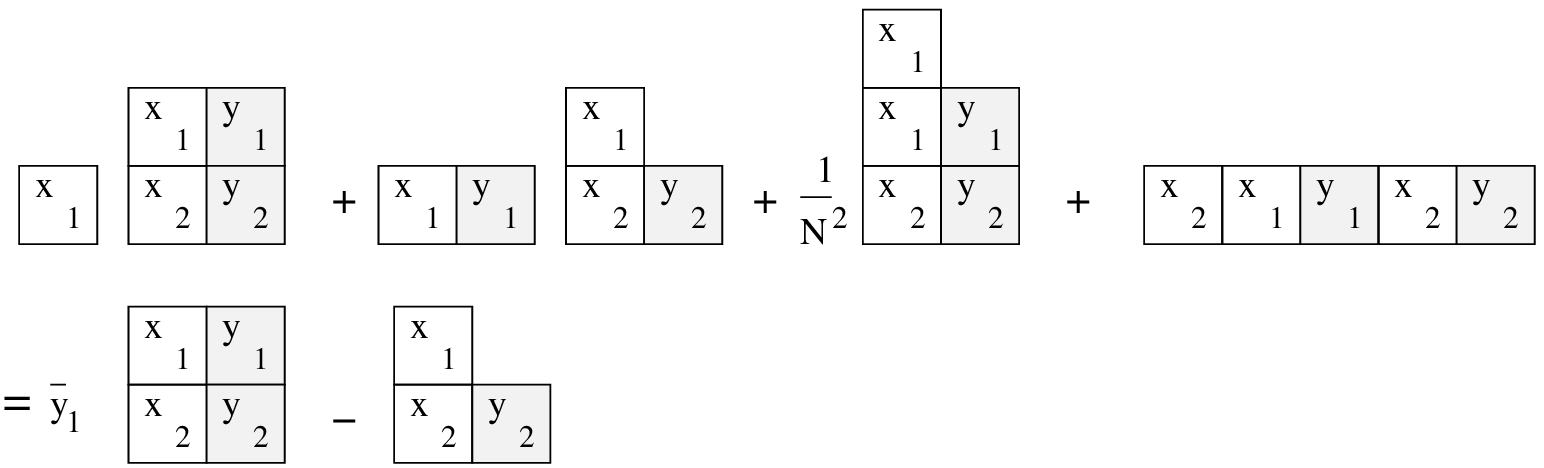}} 
\end{center}
\end{figure}
\vspace*{-2.3cm}

\noindent
\bea\label{W11loop}
\lefteqn{W_{1,1}(x_1,y_1;x_2,y_2)(\yb_1-\frac{1}{x_1}) }\\
&&=\left(1+W_1(x_1,y_1)\right) U_1(x_1;x_2,y_2)
+\frac{W_2(x_1,y_1,x_2,y_2)-W_2(x_2,y_1,x_2,y_2)}{x_2-x_1}
\nonumber \\
&&{}+\frac{1}{N^2}U_{1,1}(x_1;x_1,y_1;x_2,y_2) \nonumber
\eea
Now, making use of~\rf{W10},~\rf{W2leading} and~\rf{U10} we get
\beq\label{W110}
W_{1,1}^{(0)}(x_1,y_1;x_2,y_2) =  {x_1x_2 \yb_1\yb_2 (1+ x_1x_2\yb_1\yb_2
(x_1\yb_2+x_2\yb_1-3))\over
(x_1\yb_1-1)^2(x_1\yb_2-1)^2(x_2\yb_1-1)^2(x_2\yb_2-1)^2}
\eeq
where we note that
\beq
(1+ x_1x_2\yb_1\yb_2(x_1\yb_2+x_2\yb_1-3)) =
\det \pmatrix{1 & x_1\yb_2 & x_1\yb_2 \cr x_2\yb_1 & 1 & 
x_2\yb_1 \cr x_2\yb_1 & x_1\yb_2 & 1 }
\eeq
From~\rf{W110} we can easily get the genus one correction to the tree-level
mixing matrix of BMN operators coming from connected diagrams, namely
(cf.\ eqn~\rf{mixmatrix}) 
\bea\label{contour}
C_{n,m}^{(0)}&=&\sum_{p,q=0}^J 
\left< \tr (Z^{J-p} \Zbar^{J-q}) \tr (Z^p \Zbar^q) \right>_{conn}
e^{2\pi i(n p -m q)/J}\\
&=&\oint\frac{d r}{2 \pi i}\, r^{J+1}W_{1,1}(\sqrt{r}e^{-i \pi n/J},
\sqrt{r}e^{-i \pi m/J};\sqrt{r}e^{i \pi n/J},\sqrt{r}e^{i \pi m/J})\,
e^{i \pi (n-m)}
\nonumber \\
&=&\oint\frac{d r}{2 \pi i}
\frac{r^{J+3}(1+r^2[ 2 r \cos\left(\frac{(n+m)\pi}{J}\right)-3])\,
e^{i \pi (n-m)}}
{(r-e^{-i\pi(n-m)/J})^2(r-e^{-i\pi(n+m)/J})^2
(r-e^{i\pi(n-m)/J})^2(r-e^{i\pi(n+m)/J})^2}\nonumber 
\eea
It is obvious that the analyticity structure of the integrand
depends on the values of $n$ and $m$, more precisely we have
\begin{itemize}
\item
$n=m=0$: A pole of order 8 at $r=1$
\item
$n=0$ and $m\neq 0$ (or $m=0$ and $n\neq 0$):
Two poles of order 4 at $r=e^{\pm i \pi n/J}$ (or at $r=e^{\pm i \pi m/J}$)
\item
$n=m\neq 0$ or $n=-m\neq 0$: One pole of order 4 at $r=1$ and 2 poles of
order 2 at $r=e^{\pm 2 i\pi n/J}$· Notice that the residues are not
the same in the two cases.
\item
$|n|\neq |m|$ and $n\neq 0\neq m$: Four poles of order two at
$r=e^{\pm i \pi (n-m)/J}$ and $r=e^{\pm i \pi (n+m)/J}$
\end{itemize}
Strictly speaking, the conditions on $n$ and $m$ are to be understood 
modulo $J$ but we always consider $n$, $m \ll J$.
The fact that the evaluation of the mixing matrix has to be split
into 5 separate cases follows immediately in the generating functional
picture and evaluating the contour integral~\rf{contour} we easily reproduce
the result of reference~\cite{Kristjansen:2002bb} i.e.\
\beq
C_{n,m}^{(0)}=C_{m,n}^{(0)}= J^5\, 
\cases{\frac{1}{40} & $n=m=0$\cr
\frac{3-2\pi^2\, n^2}{24\, n^4\pi^4}& $n\neq0,m= 0$ \cr
\frac{21-2\pi^2\,n^2}{48\,n^4\pi^4}& $n=m$ and $n\neq0$
\cr
\frac{9}{32 n^4 \pi^4}&  $n=-m$ and $n\neq0$ \cr
\frac{2\, n{}^2-3\,n\,m+2\, m{}^2}
{8\,n{}^2\, m{}^2\,(n-m)^2\,\pi^4}& $|n|\neq |m|$ and 
$n\neq0\neq m$ \cr}
\eeq
up to terms of order $J^4$.
\subsection{$W_1(x,y)$ to next to leading order}

Inserting the genus expansion~\rf{genusexp1} into~\rf{W1loop}
we can easily determine the genus one contribution to
$W_1$. 
 From~\rf{W1loop} and~\rf{U10} we get
\beq
W_1^{(1)}(x,y)=\frac{x}{x\yb-1}U_1^{(0)}(x;x,y)=\frac{x^2 \yb^2}{(x\yb-1)^5}
\eeq
\section{Functionals needed for the one-loop, genus two computation}
In this section we shall determine the third order contribution to~\rf{W},
the next to leading order contributions
to the functionals~\rf{W2}--\rf{W11} as well as the leading order
contribution to the functions $U_{1,1}$, 
$W_{1,2}$, $U_2$ and $W_3$. We shall start by the latter ones
and work our way toward the first ones.

\subsection{$W_3$, $U_2$, $W_{2,1}$ and $U_{1,1}$ to leading order}

Considering the field redefinition
$\delta Z = {1\over x_1-Z}{1\over \yb_1-\Zb}{1\over x_2-Z}{1\over \yb_2-\Zb}{1\over x_3-Z}{1\over \yb_3-\Zb} $
we obtain\newpage

\begin{figure}[h]
\centerline{\epsfxsize=14.0cm\epsfbox{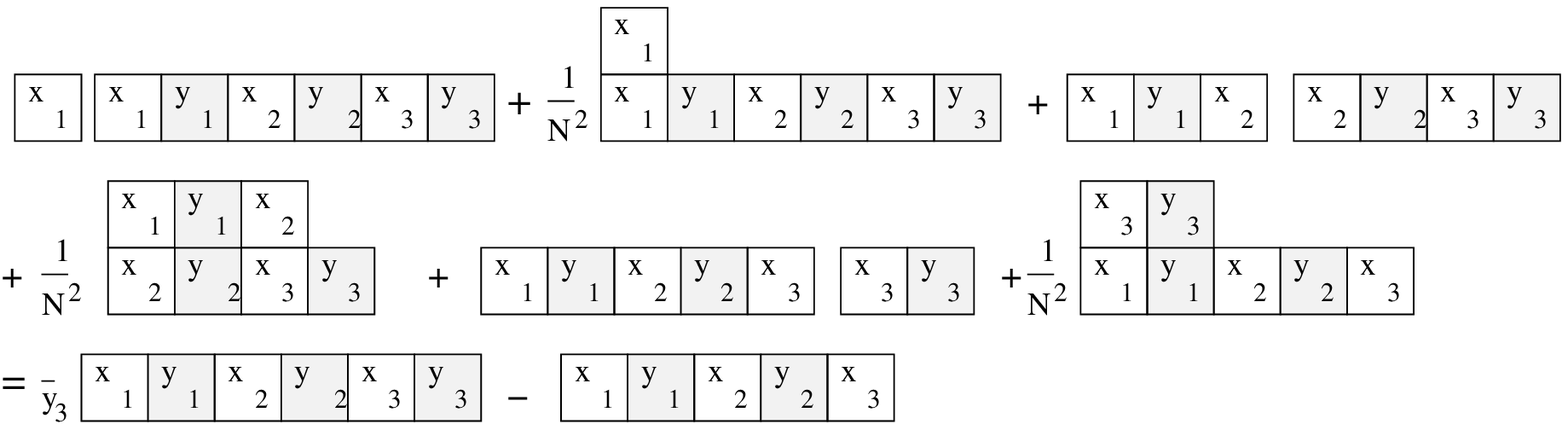}} 
\end{figure}
\vspace*{-0.6cm}

\bea\label{loopeqW3}
\lefteqn{ W_3(x_1,y_1,x_2,y_2,x_3,y_3)(\yb_3-\frac{1}{x_1})}\\
 &&= 
(1+W_1(x_3,y_3)){W_2(x_1,y_1,x_2,y_2)-W_2(x_3,y_1,x_2,y_2)\over x_3-x_1}
\nonumber \\
&&{} + W_2(x_2,y_2,x_3,y_3){W_1(x_1,y_1)-W_1(x_2,y_1)\over x_2-x_1} 
\nonumber \\
&&{}+\frac{1}{N^2}\frac{W_{1,2}(x_1,y_1;x_2,y_2,x_3,y_3)
-W_{1,2}(x_2,y_1;x_2,y_2,x_3,y_3)}{x_2-x_1} \nonumber \\
&&{}+\frac{1}{N^2}\frac{W_{1,2}(x_3,y_3;x_1,y_1,x_2,y_2)
-W_{1,2}(x_3,y_3;x_3,y_1,x_2,y_2)}{x_3-x_1} \nonumber \\
&&{}+\frac{1}{N^2}U_3(x_1;x_1,y_1,x_2,y_2,x_3,y_3)\nonumber 
\eea
From here we can immediately get the genus zero contribution to $W_3$ from
our knowledge of the genus zero contribution to $W_1$ and 
$W_2$. The result reads

\beq\label{W3leading}
W_3^{(0)}(x_1,y_1,x_2,y_2,x_3,y_3) = 
{\prod_i x_i \yb_i \over \prod_{i,j}(x_i\yb_j-1)}
\det{\pmatrix{1 & x_1\yb_2 & 1\cr 1 & 1 & x_2 \yb_3 \cr
x_3\yb_1 & 1 & 1}} 
\eeq
Next, performing the change of variables $\delta \Zb = {1\over x-Z}
\tr{1\over x_1-Z}{1\over \yb_1-\Zb}{1\over x_2-Z}{1\over \yb_2-\Zb}$ leads
to the following simple relation 
\vspace*{0.1cm}
\begin{figure}[h]
\centerline{{\epsfxsize=14.0cm\epsfbox{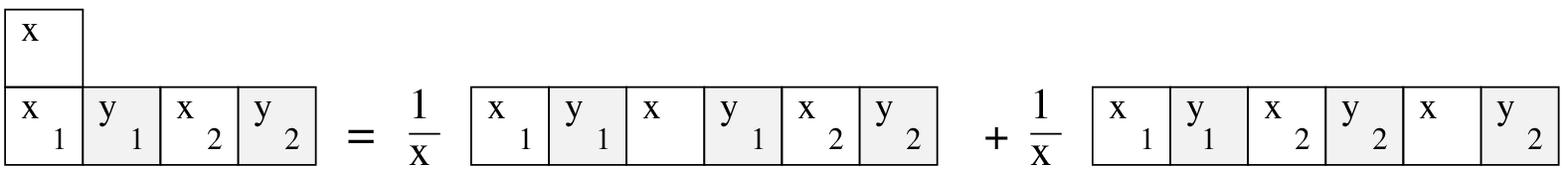}}} 
\end{figure}
\vspace*{-0.6cm}

\beq\label{loopeqU2}
 U_2(x;x_1,y_1,x_2,y_2) =\frac{1}{x}\left(
W_3(x_1,y_1,x,y_1,x_2,y_2)+W_3(x_1,y_1,x_2,y_2,x,y_2)\right)
\eeq
and from which we easily get $U_2^{(0)}$ by inserting~\rf{W3leading}.
Furthermore, choosing the field 
redefinition $\delta Z = {1\over x_1-Z}{1\over \yb_1-\Zb}
\tr {1\over x_2-Z}{1\over \yb_2-\Zb}{1\over x_3-Z}{1\over \yb_3-\Zb}$ we find
\newpage

\begin{figure}[h]
\centerline{{\epsfxsize=14.0cm\epsfbox{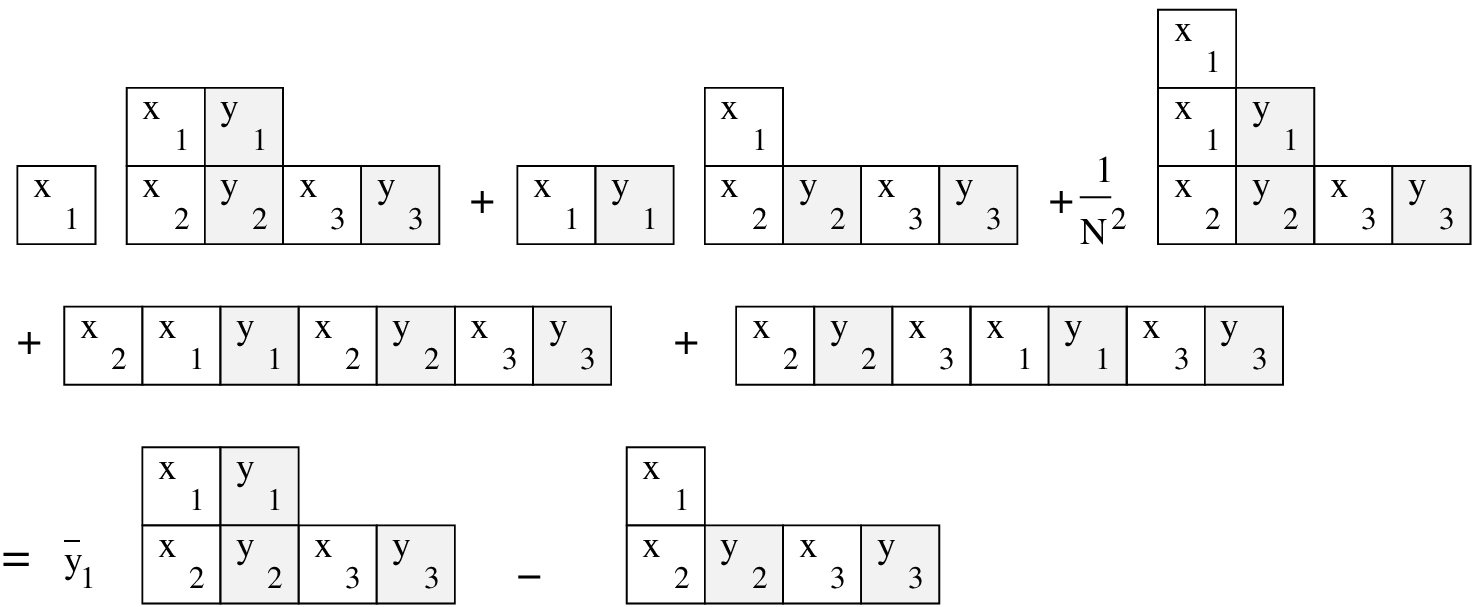}}} 
\end{figure}
\vspace*{-0.2cm}

\bea\label{loopeqW21}
\lefteqn{W_{1,2}(x_1,y_1;x_2,y_2,x_3,y_3)(\yb_1-{1\over x_1}) }\\
& &=  (1+W_1(x_1,y_1))U_2(x_1;x_2,y_2,x_3,y_3) \nonumber \\
&&{} + {W_3(x_2,y_1,x_2,y_2,x_3,y_3)-W_3(x_1,y_1,x_2,y_2,x_3,y_3)
\over x_1-x_2} 
\nonumber \\
&&{} + {W_3(x_3,y_3,x_2,y_2,x_3,y_1)-W_3(x_3,y_3,x_2,y_2,x_1,y_1)
\over x_1-x_3}
\nonumber \\
&&{}+\frac{1}{N^2}U_{1,2}(x_1;x_1,y_1;x_2,y_2,x_3,y_3)
\nonumber
\eea
where $W_{1,2}$ is now expressed in terms of already known quantities.
Furthermore, making use of the change of variable $\delta{\Zb} = 
{1\over x-Z}\tr {1\over x_1-Z}{1\over \yb_1-\bar{Z}} 
\tr {1\over x_2-Z}{1\over \yb_2-\bar{Z}}$
we obtain the following simple relation

\begin{figure}[h]
\centerline{\epsfbox{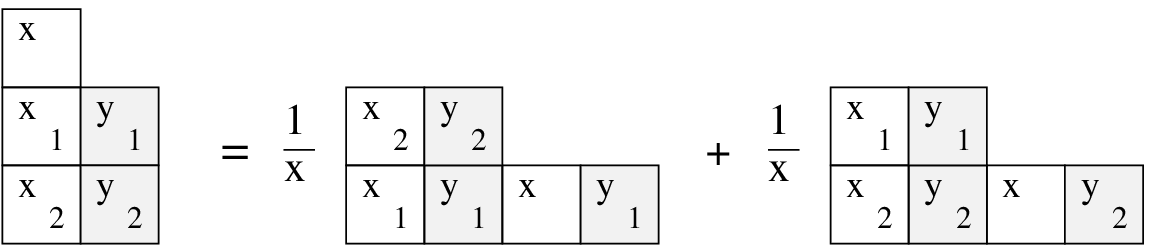}} 
\vspace*{-.8cm}
\end{figure}

\beq\label{loopeqU11}
 U_{1,1}(x;x_1,y_1;x_2,y_2) = \frac{1}{x}\left(
W_{1,2}(x_2,y_2;x_1,y_1,x,y_1)
+ W_{1,2}(x_1,y_1;x_2,y_2,x,y_2)\right)
\eeq

\subsection{$W_2$, $U_1$ and
$W_{1,1}$ to next to leading order}

Inserting the genus expansion~\rf{genusexp1} and~\rf{genusexp2} into the
relevant loop equations~\rf{W2loop}, ~\rf{loopeqU1} 
and~\rf{W11loop} we can easily determine the genus one contribution to
$U_1$, $W_2$ and 
$W_{1,1}$.
From~\rf{W2loop} we find (using Mathematica)
\bea
\lefteqn{ W_2^{(1)}(x_1,y_1,x_2,y_2)  = }\\
&& \frac{x_1}{x_1\yb_2-1}
\left\{(1+W_1^{(0)}(x_2,y_2))
{W_1^{(1)}(x_1,y_1)-W_1^{(1)}(x_2,y_1)\over x_2-x_1}\right.
\nonumber \\
&&\left.
+{W_1^{(0)}(x_1,y_1)-W_1^{(0)}(x_2,y_1)\over x_2-x_1}W_1^{(1)}(x_2,y_2) 
\right.
\nonumber \\
&&\left. +{W_{1,1}^{(0)}(x_1,y_1;x_2,y_2)-W_{1,1}^{(0)}(x_2,y_1;x_2,y_2)
\over x_2-x_1} + U_2^{(0)}(x_1;x_1,y_1,x_2,y_2) \right\} \nonumber \\
&&=\frac{\prod_i x_i\yb_i}{\prod_{i,j} (x_i\yb_j-1)^5} 
\mbox{Pol}_{28}(x_1,\yb_1,x_2,\yb_2) \nonumber
\eea
where $\mbox{Pol}_d(x_1,\yb_1,x_2,\yb_2)$ denotes a polynomial of degree
$d$ in $x_1$, $\yb_1$, $x_2$, $\yb_2$.
If we carry out the discrete Fourier transform and take the large-$J$ limit
we get
\bea
\lefteqn{
D_{n,m}^{(1)}=D_{m,n}^{(1)}=\sum_{p,q=0}^J
\frac{1}{N}\left<\tr Z^p \Zb^q Z^{J-p} \Zb^{J-q}\right>_{h=1}
e^{\frac{2 \pi i}{J}(np-mq)} } \\
&=&
\oint\frac{d r}{2 \pi i}\, r^{J+1}W_2^{(1)}(\sqrt{r}e^{-i \pi n/J},
\sqrt{r}e^{-i \pi m/J};\sqrt{r}e^{i \pi n/J},\sqrt{r}e^{i \pi m/J})\,
e^{i \pi (n-m)}\nonumber \\
&=&
J^7\, 
\cases{\frac{1}{240} & $n=m=0$\cr
\frac{6-\pi^2\, n^2}{48\, n^4\pi^4}& $n\neq0,m= 0$ \cr
\frac{315-120\pi^2\,n^2+32 \pi^4 n^4}{7680\,n^6\pi^6}& 
$|n|=|m|$ and $n\neq0$
\cr
\frac{3(n^6+m^6)+4\pi^2(n^6 m^2+m^6 n^2)-15(m^4 n^2+n^4 m^2)-8m^4n^4 \pi^2}
{48\,n{}^2\, m{}^2\,(m-n)^4\,(m+n)^4\pi^6}& $|n|\neq |m|$ and 
$n\neq0\neq m$ \cr} \nonumber 
\eea
up to terms of order $J^6$.

Furthermore, from~\rf{loopeqU1}
\bea\label{U1one}
\lefteqn{U_1^{(1)}(x;x_1,y_1) = {1\over x}W_2^{(1)}(x,y_1,x_1,y_1)=} \\
&&\frac{x_1 \yb_1^2}{(1-x \yb_1)^6(1-x_1 \yb_1)^6}
 \left\{1+
4\yb_1 (x_1+x)
-34 \yb_1^2 x x_1 \right.\nonumber \\
&&
18 \yb_1^3(x x_1 (x+x_1))
+ \yb_1^4 x x_1 (-10 x^2 + 21 x x_1 -10 x_1^2) \nonumber \\
&&\left.+ 2 \yb_1^5 x x_1 ( x^3 - 6 x^2 x_1 - 6 x x_1^2 + x_1^3)
+ \yb_1^6 x^2 x_1^2 (3 x^2 + 2 x x_1 + 3 x_1^2)
\right\} \nonumber
\nonumber 
\eea
Finally, from~\rf{W11loop} one gets
\bea
\lefteqn{W_{1;1}^{(1)}(x_1,y_1;x_2,y_2)=} \\
& &  \frac{x_1}{x_1 \yb_1-1}\left\{
(1+W_1^{(0)}(x_1,y_1)) U_1^{(1)}(x_1;x_2,y_2) +
W_1^{(1)}(x_1,y_1) U_1^{(0)}(x_1;x_2,y_2) \right. \cr
&& \left.+ {W_2^{(1)}(x_1,y_1,x_2,y_2)-W_2^{(1)}(x_2,y_1,x_2,y_2)\over x_2-x_1}
+ U_{1,1}^{(0)}(x_1;x_1,y_1;x_2,y_2)\right\} \cr
&&=\frac{\prod_i x_i\yb_i}{\prod_{i,j} (x_i\yb_j-1)^6}
\mbox{Pol}_{36}(x_1,\yb_1,x_2,\yb_2) \nonumber
\eea
In this case carrying out the discrete Fourier transform and taking the
large-$J$ limit gives
$$
C_{n,m}^{(1)}=C_{m,n}^{(1)}= J^9\, 
\cases{\frac{73}{181440} & $n=m=0$\cr
\frac{7}{16n^8 \pi^8}-\frac{13}{96n^6 \pi^6}+\frac{7}{320n^4 \pi^4}
-\frac{1}{480 n^2 \pi^2} 
& $n\neq0$ and $m=0$
\cr
\frac{143}{2048n^8\pi^8}-\frac{107}{768 n^6 \pi^6}
+\frac{3}{160 n^4 \pi^4}-\frac{1}{2016 n^2\pi^2}& $n=m$ and $n\neq0$ \cr
\frac{7}{4096 n^8\pi^8}-\frac{245}{3072 n^6\pi^6}+\frac{13}{1280 n^4\pi^4}&  
$n=-m$ and $n\neq0$ \cr
\frac{1}{192 m^4 n^4(m-n)^6(m+n)^4\pi^8} \times {\mbox{Pol}}_{14}(m,n)
& $|n|\neq |m|$ and 
$n\neq0\neq m$ \nonumber }   
$$
where
\bea
{\mbox{Pol}}_{14}(m,n)&=&
3\left( 2\,m^{10} - m^9\,n - 13\,m^8\,n^2 + 
       9\,m^7\,n^3 + 63\,m^6\,n^4 + 120\,m^5\,n^5 \right. \nonumber  \\
&&{}+ \left.  63\,m^4\,n^6 + 9\,m^3\,n^7 - 13\,m^2\,n^8 - m\,n^9 + 
       2\,n^{10} \right) \nonumber \\
&&{} - 
    {\left( m^2 - n^2 \right) }^2\,
     \left( 12\,m^8 - 18\,m^7\,n - 20\,m^6\,n^2 + 
       37\,m^5\,n^3 \right. \nonumber \\
&&{}\left. + 38\,m^4\,n^4 + 37\,m^3\,n^5 - 
       20\,m^2\,n^6 - 18\,m\,n^7 + 12\,n^8 \right) \,{\pi }^2
 \nonumber \\
&&{}+ m^2\,n^2\,{\left( m^2 - n^2 \right) }^4\,
     \left( 2\,m^2 - 3\,m\,n + 2\,n^2 \right) \,{\pi }^4
\eea
and where we have neglected terms of order $J^8$. 
Notice that the expressions above constitute the
contribution to the mixing matrix coming from connected diagrams only.
If one includes also disconnected ones one reproduces the expressions
given in~\cite{Constable:2002hw}.\footnote{A detailed comparison is not
possible in the general $|n|\neq |m|$ case and there seems to be  a
discrepancy in the two last terms of the $n=-m$ case.}

\subsection{$W_1(x,y)$ to third order}

Making use of~\rf{W1loop} and~\rf{U1one} we get
\beq
W_1^{(2)}(x,y)=\frac{x}{x\yb-1}U_1^{(1)}(x;x,y)=
\frac{x^2 y^2(1+12 x \yb+8x^2\yb^2)}{(x\yb-1)^9}
\eeq

\section{The general case}
From the examples above it should be clear how to choose the appropriate
field redefinitions needed for the derivation of the loop equation associated
with a given generating functional. Here we shall write down the most 
general loop equations and show how they allow us to determine recursively
any multi-loop correlator, i.e.\ any expectation value of traces of
words to any order in the genus expansion. 

\subsection{One-trace functions}
Considering the change of variable $\delta \Zb={1\over x_1-Z} 
{1\over \yb_1-\Zb}\ldots{1\over x_l-Z}{1\over \yb_l-\Zb}$ with $l\geq 2$
we obtain the following relation
\bea\label{recursion}
\lefteqn{\hspace{-1.0cm}(1/x_1-\yb_l)W_{l}(x_1,\dots,y_l) =}\\
&&   \sum_{k=2}^{l} {W_{k-1}(x_1,y_1,\dots,y_{k-1})
-W_{k-1}(x_k,y_1,\dots,y_{k-1})\over x_1-x_k} \times \cr
&&\,\,\,\,\,\,\,\, (\delta_{k,l}+W_{l-k+1}(x_k,y_k,\dots,x_l,y_{l})) 
+O(\frac{1}{N^2})
\nonumber
\eea
Here $W_l$ is expressed entirely in terms of $W_k$ with $k\leq l-1$.
Having determined the planar contribution to $W_1(x,y)$ 
(cf.\ eqn.~\rf{W10}) we can by means of~\rf{recursion} determine
recursively the planar contribution to any one-trace function $W_l$.
From the structure of the recursion relation and the explicit expression
for $W_1^{(0)}(x,y)$ it follows that planar one-trace functions only have
singularities in the form of single poles. More precisely we have
\beq
W_l^{(0)}(x_1,y_1,\ldots,x_l,y_l)=
\frac{\prod_i x_i \yb_i}{\prod_{i,j}(x_i \yb_j-1)} 
\mbox{Pol}(x_1,\yb_1,\ldots,x_l,\yb_l)
\eeq
where Pol$(x_1,\yb_1,\ldots,x_l,\yb_l)$ is a polynomial 
of degree $l-2$ in each of its variables. In the case $l=3$ 
(and trivially in the case $l=2$) this polynomial could be expressed
as a determinant (cf.\ equations~\rf{W2leading} and~\rf{W3leading})
 but it does not seem that a similar simplification
occurs for higher values of $l$. It would be most interesting, though,
to find a closed expression for $W_l$ for general $l$.

\subsection{Multi-trace functions}

In the case of the $W$-functions
it is convenient to consider separately the cases $l_1=1$ and $l_1>1$.
For $l_1=1$ and ($n\geq2$) we have
\bea\label{leq1}
\lefteqn{\hspace{-1.0cm} (\yb_{1,1}-{1\over x_{1,1}}) W_{1,l_2,\dots,l_n} (x_{1,1},y_{1,1};
x_{2,1},\dots,y_{n,l_n})=} \\
&&  (1+W(x_{1,1},y_{1,1}) ) U_{l_2,\dots,l_n} (x_{1,1};x_{2,1},\dots,
y_{n,l_n})  \cr
&&+ \sum_{k=2}^n \sum_{j=1}^{l_k}{1\over x_{k,j}-x_{1,1}}\left\{
W_{l_2,\dots,l_k+1,\dots,l_n}(\dots,y_{k,j-1},x_{1,1},y_{1,1},x_{k,j},
y_{k,j},\dots) \right. \cr
&&\left.
-W_{l_1,\dots,l_k+1,\dots,l_n}(\dots,y_{k,j-1},x_{k,j},y_{1,1},x_{k,j},
y_{k,j},\dots) \right\} \cr
&& + {1\over N^2} U_{1,l_2,\dots,l_n} (x_{1,1};x_{1,1},y_{1,1};x_{2,1},
\dots,y_{n,l_n}) \nonumber
\eea
whereas for $l_1>1$ the relevant loop equation reads
\bea\label{lgt1}
\lefteqn{\hspace{-.5cm}(\yb_{1,l_1}-{1\over x_{1,1}}) 
 W_{l_1,..,l_n}(x_{1,1},..,y_{n,l_n}) }
 \\
&&=\sum_{j=2}^{l_1} W_{l_1+1-j,..,l_n}(x_{1,j},..,y_{n,l_n})(1-\delta_{n,1})
\times \cr 
&&\left\{{W_{j-1}(x_{1,1},y_{1,1},..,y_{1,j-1})-
W_{j-1}(x_{1,j},y_{1,1},..,y_{1,j-1})\over x_{1,j}-x_{1,1}}\right\}  \cr
&& + \sum_{j=2}^{l_1}  (W_{l_1+1-j}(x_{1,j},..,y_{1,l_1})+\delta_{j,l_1}) 
\times \cr 
&&\left\{
{W_{j-1,..,l_n}(x_{1,1},..,y_{1,j-1};x_{2,1},..,y_{n,l_n})-W_{j-1,..,l_n}
(x_{1,j},..,y_{1,j-1};x_{2,1},..,y_{n,l_n})\over x_{1,j}-x_{1,1}}
\right\} \cr
&& + \sum_{k=2}^n \sum_{j=1}^{l_k} {1\over x_{k,j}-x_{1,1}}
\left\{
W_{l_2,..,l_k+l_1,..,l_n}(x_{2,1},..,y_{k,j-1},x_{1,1},..,y_{1,l_1},x_{k,j},..,y_{n,l_n}) \right. \cr
&&
\left.-W_{l_1,..,l_k+l_1,..,l_n}(x_{2,1},..,y_{k,j-1},x_{k,j},..,y_{1,l_1},x_{k,j},..,y_{n,l_n})\right\} \cr
&& + {1\over N^2} 
\sum_{j=2}^{l_1} {1 \over x_{1,j}-x_{1,1}}\left\{
W_{j-1,l_1+1-j,l_2,..,l_n}(x_{1,1},y_{1,1},..,y_{1,j};x_{1,j},..,y_{n,l_n})
\right. \cr
&&\left.
-W_{j-1,l_1+1-j,..,l_n}(x_{1,j},y_{1,1},..,y_{1,j};x_{1,j},..,y_{n,l_n}) 
\right\}\cr
&&+{1\over N^2}U_{l_1\ldots l_n}(x_{1,1};x_{1,1},\ldots y_{n,l_n})
 \nonumber
\eea
These equations are to be supplemented by the loop equations
for the $U$-functions which take the simpler form
\bea\label{Ugeneral}
\lefteqn{ x\,\, U_{l_1,\dots,l_n}(x;x_{1,1},\dots,y_{n,l_n})} \\
&& = \sum_{k=1}^n \sum_{j=1}^{l_k}
W_{l_1,\dots,l_k+1,\dots,l_n} (x_{1,1},\dots;x_{k,1},\dots,y_{k,j},x,y_{k,j},x_{k,j+1},\dots,y_{k,l_k};\dots,y_{n,l_n}) 
\nonumber
\eea
The relations~\rf{leq1}, \rf{lgt1} and~\rf{Ugeneral} constitute a
triangular set of equations which allows us to determine any multi-loop
function to any order in the genus expansion. For a finite number of
loops and finite genus only a finite number of operations are needed.
Below we shall make this statement more precise.

First, let us introduce an ordering of the Young
diagrams representing $W$-functions. 
A Young diagrams $Y_{k_1,\ldots,k_m}$ is said to be smaller than a Young
diagram $Y_{l_1,\ldots,l_n}$ 
(representing loop-functions $W_{k_1,\ldots,k_m}$ and $W_{l_1,\ldots,l_n}$,
respectively)
if:
\beq
 m+\sum_{i=1}^{m} k_i  < n+\sum_{i=1}^n l_i 
\eeq
Next, let us consider the loop equations~\rf{leq1} and~\rf{lgt1} 
(using~\rf{Ugeneral}):
it is clear that all the leading order diagrams on the RHS are smaller
than the diagram on the LHS.
 This means that at the planar level a $W$-function corresponding to a
certain Young diagram can be
expressed entirely in terms of planar $W$-functions corresponding to
smaller
Young diagrams. We thus get a closed equation for any genus zero
$W$-function and clearly also for any genus zero $U$-function (cf.\
equation~\rf{Ugeneral}). 

Proceeding to higher genera, we have in our
loop equations~\rf{leq1} and~\rf{lgt1} two types of terms which carry
a factor $1/N^2$; $U$-terms and $W$-terms. Compared to the object we
are interested in, the $W$-terms correspond to Young diagrams where
one extra line has been added while the number of boxes has been 
kept fixed. The $U$-terms, on the other hand, correspond
via~\rf{Ugeneral} to Young diagrams where two extra boxes have been 
added while the number of lines has been kept fixed. This means that 
the genus $g$ contribution to a $W$-function described by a Young 
diagram with $2K$ boxes and $n$ lines can be expressed entirely in
terms of genus zero $W$-functions corresponding to Young diagrams
having at most $2K+2g$ boxes and $n+g-1$ lines. Clearly, we have a triangular
set of equations which allows us to determine any expectation value
of traces of words to any order in the genus expansion.

\section{Exactly calculable correlators}
As shown in reference~\cite{Beisert:2002bb,Okuyama:2002zn} it is
possible to find exact expressions for the expectation values encoded
in the following generating functionals
\beq\label{H}
H_n(x_1,\ldots,x_n,y)=N^{n-1} \left< \tr{1 \over x_1-Z}\ldots
\tr{1 \over x_n-Z} \tr{1\over \yb-\Zb} \right >_{conn}
\eeq
as well as~\cite{Beisert:2002bb}
\beq\label{G}
G(x_1,x_2,y_1,y_2)=N^2\left<\tr {1 \over x_1-Z}\tr {1\over x_2-Z}
\tr {1 \over \yb_1-\Zb}\tr {1\over \yb_2-\Zb}\right >_{conn}
\eeq
From the generating functionals~\rf{H} and~\rf{G}
it is possible using again loop
equations
to derive exact expressions for yet other generating functionals. The
functions $W_1(x,y)$ and $U_1(x';x,y)$ can be determined in full
generality whereas the method only gives the remaining $W$- and
$U$-functions in certain limits where typically a number of their
arguments are sent to $\infty$. 
To obtain the all genus version of $W_1(x,y)$ one considers the field
redefinition $\delta \Zb= {1\over x-Z}\tr {1 \over \yb-\Zb}$ which leads
to the following first order differential equation
\beq\label{diffeqn1}
-\partial_\yb W_1(x,y)=x H_1(x,y)
\eeq
which is to be supplemented by the boundary condition
\beq
W_1(x,y)\rightarrow \frac{1}{x\yb}, \hspace{0.5cm}\mbox{as}
\hspace{0.5cm} |x|, |y|\rightarrow \infty
\eeq
The equation~\rf{diffeqn1} is of course nothing but the generating
functional version of the simple relation
\beq
\left< \tr Z^J\tr \Zb^J\right> = J\left< \tr Z^{J-1}\Zb^{J-1}\right>
\eeq
which implies
\beq
W_1(x,y) = \sum_{J=0}^\infty {1\over N^J x^{J+1} \yb^{J+1}} 
{1\over (J+1)(J+2)} \left\{{(N+J+1)!\over (N-1)!}-{N!\over (N-J-2)!}\right\}
\eeq
which can also be written
\beq
W_1(x,y) =  \sum_{k=0}^\infty {1\over N^{2k}} f_{2k+1}(x\yb)
\eeq
with $f_1(x)=1/(x-1)$ and:
\beq
{\D^2\over \D{x}^2}\, f_{k+1} = {1\over 1-x}\,\, {\D\over \D{x}}\, x\, {\D^2\over \D{x}^2} \, f_{k}
\eeq
and $f_k(x) = O(1/x^{k})$ for large $x$.
The coefficient of $1/(x-1)^{2k-1}$ as $x\rightarrow 1$  is $(2k-3)!!/k$.

In the case of $U_1(x';x,y)$ one chooses the field redefinition
$\delta \Zb= {1\over x-Z}\tr {1 \over x'-Z}
\tr {1 \over \yb-\Zb}$ and obtains
\beq\label{H2}
-\partial_\yb U_1(x';x,y)=x H_2(x',x,y) 
\eeq
and the appropriate boundary condition in this case reads
\beq
U_1(x';x,y)\rightarrow \frac{1}{(x'\yb)^2 x }\hspace{0.5cm}\mbox{as}
\hspace{0.5cm} |x'|, |x|, |y|\rightarrow \infty
\eeq
Expressed in terms of expectation values~\rf{H2} reads
\beq
\left<\tr Z^J \tr Z^K \tr \Zb^{J+K}\right>=(J+K)
\left< \tr Z^K \tr \Zb^{J+K-1} Z^{J-1}\right>
\eeq
which has the obvious generalization with $J=\sum_{i=1}^k J_i$
\beq
\left< \tr \Zb^J \prod_{i=1}^k \tr Z^{J_i}\right>
=
J \left< \tr \Zb^{J-1}Z^{J_1-1} \prod_{i=2}^k \tr Z^{J_i} \right>
\eeq
\section{Conclusion}
With this work we have added pp-wave physics and ${\cal N}=4$ SYM to the
long list of areas where classical matrix model techniques have proven 
very efficient.

As explained in section~2, evaluating a typical correlation function in
the BMN sector of ${\cal N}=4$ SYM can be reduced to evaluating the expectation
value of a product of traces of words in a zero-dimensional Gaussian
complex matrix model and subsequently carrying out a discrete Fourier
transformation. Simple correlators can be obtained by purely combinatorial
arguments but such arguments become more and more involved (and
correspondingly less and less reliable) the more words enter the
correlators and the higher genus one is aiming at. With our loop
equation based technique, however, we can by analytical manipulations
reach any multi-word correlator to any order in the genus expansion. 
Furthermore, by working with generating functionals we trade the
process of Fourier transformation for simple contour integration.

There are several directions of investigation where our technique
would be most useful. One is the investigation of operators with more
impurities than the traditionally studied BMN operators of 
equation~\rf{BMNop}. Such operators would correspond to string states
with many oscillators excited and determining their correlators would
imply evaluating expectation values of many letter words. Such words
are encoded in $U$- and $W$-functions whose sub-scripts are large and
these are of course accessible with our method.

As mentioned several times our method would also allow us to calculate
the higher genera, one-loop corrections to the anomalous dimension of 
the BMN operators~\rf{BMNop}. In fact, we have already evaluated all
expectation values needed for the genus two calculation. As pointed
out in~\cite{Beisert:2002bb}, completing this calculation would allow 
one to check whether the effective string coupling constant in the
pp-wave/BMN correspondence is indeed $g_2\sqrt{\lambda'}$ as suggested
in~\cite{Constable:2002hw,Verlinde:2002ig}. Finally, it is possible
that like in the one-loop case two- and higher loop computations on
the gauge theory side can be reduced to pure matrix model computations
and then obviously our method will again be in demand.

\end{document}